\documentclass[referee]{aa} 
\usepackage{graphicx}
\usepackage{natbib} 
\bibpunct{(}{)}{;}{a}{}{,} 

\newcommand{\solm}{M$_{\odot}$}

\newcommand{\phn}{\phantom{0}}
\newcommand{\phnn}{\phantom{00}}
\newcommand\arcdeg{\mbox{$^\circ$}}%

\usepackage{txfonts}
\begin{document}
   \title{First Simultaneous NIR/X-ray Detection of a Flare from SgrA*}


   \author{A. Eckart\inst{1}
          \and
          F. K. Baganoff\inst{2}
          \and 
          M. Morris\inst{3}
          \and
          M.W. Bautz\inst{2}
          \and
          W.N. Brandt\inst{4}
          \and
          G.P. Garmire\inst{4}
          \and
          R. Genzel\inst{5,6}
          \and
          ~T. Ott\inst{5}
          \and
          ~G.R. Ricker\inst{2}
          \and
          ~C. Straubmeier\inst{1}
          \and
          ~T. Viehmann\inst{1}
          \and
          ~R. Sch\"odel\inst{1}
          \and
          ~G.C. Bower\inst{6}
          \and
          J.E. Goldston\inst{6}
          }

   \offprints{A. Eckart}

   \institute{I.Physikalisches Institut, Universit\"at zu K\"oln,
              Z\"ulpicher Str.77, 50937 K\"oln, Germany\\
              \email{eckart@ph1.uni-koeln.de}
         \and
             Center for Space Research, Massachusetts Institute of
              Technology, Cambridge, MA~02139-4307, USA\\
             \email{fkb@space.mit.edu}
         \and
             Department of Physics and Astrometry, University of 
             California Los Angeles, Los Angeles, CA~90095-1562, USA
         \and
             Department of Astronomy and Astrophysics, Pennsylvania
              State University, University Park, PA~16802-6305, USA
         \and
             Max Planck Institut f\"ur extraterrestrische Physik,
              Giessenbachstra{\ss}e, 85748 Garching, Germany
         \and
             Department of Astronomy and Radio Astronomy Laboratory,
             University of California at Berkeley, 601 Campbell Hall, 
             Berkeley, CA~94720, USA
             }

   \date{Received ; Accepted }

   \abstract{We report on the first simultaneous near-infrared/X-ray
detection of the SgrA* counterpart which is associated with the
massive 3--4$\times$10$^6$\solm ~black hole at the center of the Milky Way.  
The observations have been carried out using the NACO adaptive
optics (AO) instrument at the European Southern Observatory's Very Large
Telescope\footnote{Based on observations at the Very Large Telescope
(VLT) of the European Southern Observatory (ESO) on Paranal in Chile;
Program: 271.B-5019(A).} and the ACIS-I instrument aboard the
\emph{Chandra X-ray Observatory}.  
We also report on quasi-simultaneous observations at a wavelength of 3.4~mm
using the Berkeley-Illinois-Maryland Association (BIMA)
array.
A flare was detected in the
X-domain with an excess 2 - 8 keV luminosity of about
6$\times$10$^{33}$~erg/s.  A fading flare of Sgr~A* with $>$2 times
the interim-quiescent flux was also detected at the beginning of the
NIR observations, that overlapped with the fading part of the X-ray
flare.  
Compared to 8-9 hours before the NIR/X-ray flare 
we detected a marginally significant increase in the millimeter 
flux density of Sgr A* during measurements about 7-9 hours afterwards. 
We find that the flaring state can be conveniently explained
with a synchrotron self-Compton model involving up-scattered
sub-millimeter photons from a compact source component, possibly with 
modest
bulk relativistic motion. 
The size of that component is assumed to be of the order of a few times 
the Schwarzschild radius.  The overall spectral indices
$\alpha_{NIR/X-ray}$ ($S_{\nu}$$\propto$$\nu^{-\alpha}$) of both
states are quite comparable with a value of $\sim$1.3.  Since the
interim-quiescent X-ray emission is spatially extended, the spectral index for
the interim-quiescent state is probably only a lower limit for the
compact source Sgr~A*.  A conservative estimate of the upper limit of
the time lag between the ends of the NIR and X-ray flare is of the
order of 15~minutes.   }

   \titlerunning{Simultaneous NIR/X-ray Flare from Sgr~A*}
   \authorrunning{Eckart, Baganoff, Morris et al.}  
   \maketitle

%

\section{Introduction}

Over the last decades, evidence has been accumulating that most quiet
galaxies harbor a massive black hole (MBH) at their centers.
Especially in the case of the center of our Galaxy, progress could be
made through the investigation of the dynamics of stars (Eckart \&
Genzel 1996, Genzel et al. 1997, 2000, Ghez et al. 1998, 2000, 2003a,
2003b, Eckart et al. 2002, Sch\"odel et al. 2002, 2003).  Located at a
distance of only 8 kpc from the solar system (Reid 1993, Eisenhauer et
al. 2003), it allows detailed observations of stars at distances much
less than 1~pc from the central black hole candidate, the compact radio 
source Sgr~A*.  Additional strong evidence for a massive black hole at the
position of Sgr~A* came from the observation of interim-quiescent and 
flare
activity from that position both in the
X-ray and recently in the near-infrared wavelength domain (Baganoff et
al. 2001, 2003, Eckart et al. 2003, Porquet et al. 2003, Goldwurm et
al. 2003, Genzel et al. 2003, Ghez et al. 2004).  Throughout the paper
we will use the term 'interim-quiescent' (or IQ) for the apparently
constant, low-level flux density states at any given observational
epoch since current data cannot exclude flux density variations of
that state on longer time scales (days to years). This is especially
true for the more compact NIR source (Genzel et al. 2003, Ghez et
al. 2004).

Simultaneous observations of SgrA* across different wavelength regimes
are of high value, since they provide information on the emission
mechanisms responsible for the radiation from the immediate vicinity
of the central black hole.  The first observations of SgrA* covering
an X-ray flare simultaneously in the near-infrared using seeing
limited exposures revealed only upper limits to the NIR flux density
(Eckart et al. 2003). In section 2 of the present paper we report on
the first successful simultaneous NIR/X-ray observations using
adaptive optics.  These observations were for the first time
successful in detecting radiation from the SgrA* counterpart both in
the NIR and the X-ray wavelength domain.  A detailed statistical
analysis that supports the simultaneous detection of a SgrA* flare
event in the NIR and X-ray domain is given in section 3 of this
paper. 
In section 2 we also describe the quasi-simultaneous mm-observations 
that were taken just before and after the NIR/X-ray observations.
In section 4 we briefly discuss the flux densities and spectral
indices we derived from the available data.  Section 5 gives a first
physical interpretation of the simultaneous detection of SgrA*. A
short summary and discussion of the results is given in section 6.


\section{Observations and Data Reduction}

The Galactic Center stellar cluster was observed during Directors
Discretionary Time on June~19, 2003, with the NAOS/CONICA adaptive
optics system/NIR camera at the ESO VLT unit telescope~4. The loop of
the AO was closed on the bright (K$\sim6.5$) supergiant IRS~7, located
about $6''$ north of Sgr~A*. We used the K$_{S}$-filter
($\lambda_{C}=2.18\mathrm{\mu m}$, FWHM $0.35\mathrm{\mu m}$). The
detector integration time was 10~s, with two frames added online to
final exposures of 20~s integration time. The visible seeing at zenith
was around $0.8''$.  The AO correction was stable, but of medium
quality with an estimated Strehl ratio above $20\%$ most of the
time. The correction improved during the course of the observations.
The airmasses were all less than 1.1.

The individual exposures were sky subtracted, flat-fielded and
corrected for bad pixels. We extracted PSFs for each of the images
with \emph{StarFinder} and used the extracted PSFs for a
Lucy-Richardson deconvolution of all the images. After beam
restoration with a Gaussian PSF of 60~mas FWHM, we extracted the flux
of individual sources with aperture photometry. 
The stars W6, m$_K$=14.3$\pm$0.2, W9, m$_K$=13.7$\pm$0.2, and 
W11, m$_K$=14.1$\pm$0.2, (Ott 2003) 
were used for the photometric calibration.
These stars are located only about $1'' - 2''$ west of SgrA* 
and were contained in all the images.
Under the favorable seeing conditions and small zenith angles
photometric calibration with adaptive optics over such 
angular distances can usually be applied without problems.
For precise relative photometry we used the stars 
S1, S2, and S8, all within $0.5''$ of Sgr~A* (see below).
We repeated the photometric
measurements with two different aperture sizes (26 and 39~mas) and
took the average of these measurements as the source fluxes, with the
errors given by the maximum deviation of the individual measurements
from the average.  We derived extinction corrected fluxes, assuming
A$_{\mathrm{K}}=2.8$ (Genzel et al. 2003).

\begin{figure*}
\centering
\includegraphics[width=10cm]{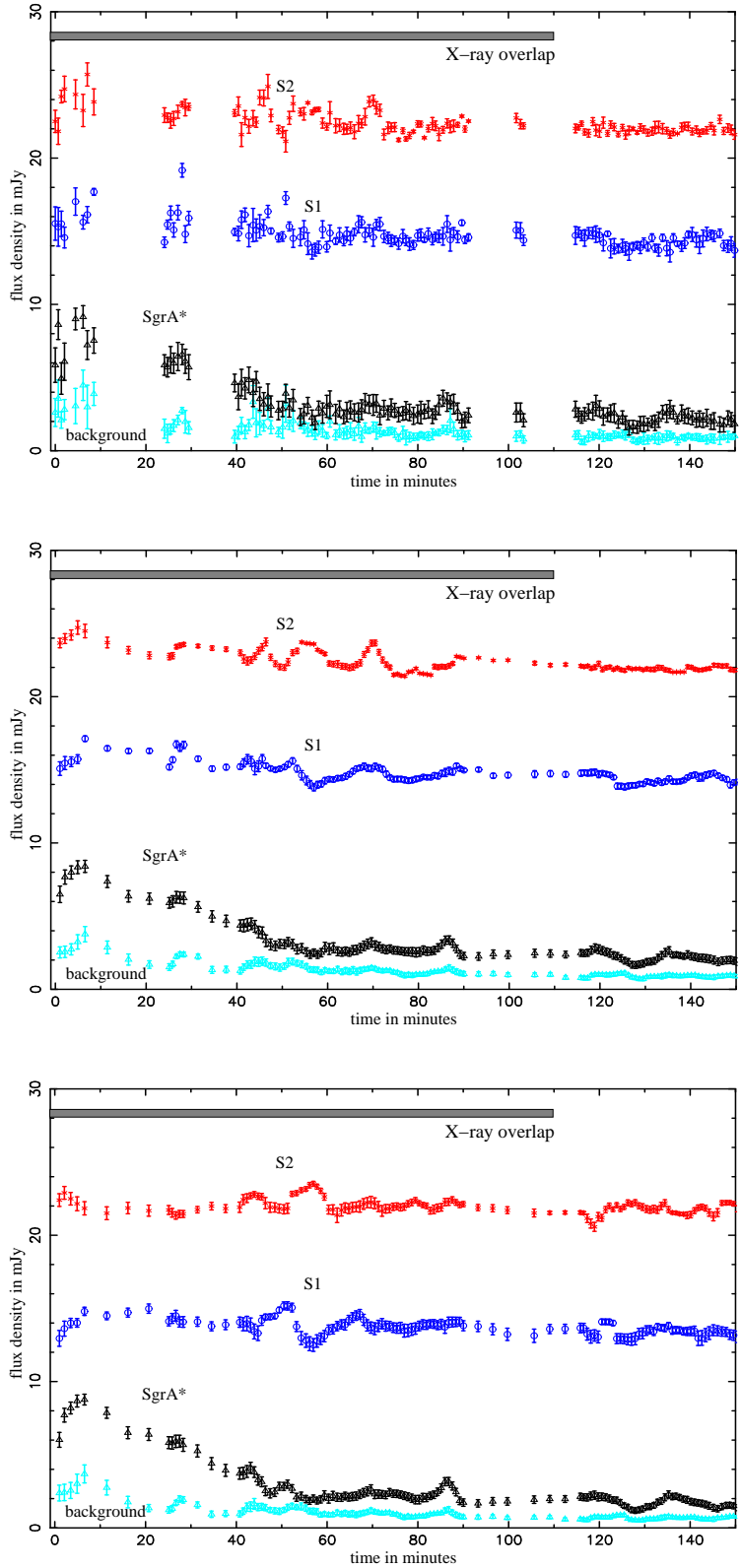}
\caption{\small NIR flux variation of Sgr~A* during the coordinated
\emph{Chandra}/VLT observations of the Galactic Center.  Upper panel:
Dereddened NIR flux vs.\ time of Sgr~A* (black), S1 (blue), S2 (red)
and of the average of random measurements in a region about $0.5''$
west of Sgr~A* with no detectable source (green). Error bars were
estimated from aperture photometry with two apertures.  The
observations started on 19 June 2003 at 23:51:15 (UT).  Middle panel:
As upper panel, but smoothed with a sliding window, averaging 4
measurements at a time. Lower panel: As middle panel, but corrected
for remnant correlated flux variations.}
\label{nircurves}
\end{figure*}

In the top panel of Figure~\ref{nircurves} we show the light curves of
Sgr~A*, and of the two stars S1 and S2 (top curves: fluxes of S1 and
S2 multiplied by a factor of $1.2$), that are located in its immediate
vicinity. In order to estimate the background, we averaged aperture
photometric measurements from six random locations within a region
with no detectable source, about $0.5''$ W of SgrA*
(Figure~\ref{nircurves}, bottom light curve). For a better
visualization of the flux variations of Sgr~A* during the first
60~min, the middle panel of Figure~\ref{nircurves} shows the same
light curves of the upper panel after smoothing them with a sliding
window that averages four measurements at a time. As can be seen in
the middle panel of Figure~\ref{nircurves}, there appear still to be
some correlated remnant flux variations. 
Therefore, we used the stars S1, S2, and S8 and could
largely remove these variations by assuming a constant flux for these
stars. 
The resulting light curves are shown in the lower panel of
Figure~\ref{nircurves}.

The gaps in the measurements are due to AO reconfiguration or sky
measurements. AO correction was poor during the first ten minutes of
the observations. This is the reason for the larger error bars and the
increased background level at the beginning of the light curves. In
comparison to the light curves of the background, S1, and S2 (that are
assumed to have a constant flux, of course), the flux of Sgr~A* is
clearly increased ($>5\sigma$ as can be seen in the lower, averaged
plot) at the beginning of the observations and decreases until it
reaches almost the background level. 

Very close to the position of Sgr~A* faint stellar sources
can be detected when one averages individual images. 
We estimate that a possible systematic positive bias of the Sgr~A* 
flux density due to these faint stars is of the order of $0.5$ to 
$1.0$~mJy. This is apparently less than the total measured IQ flux density 
($\sim$2~mJy) between 55 and 150 minutes.

\begin{figure*}
\centering
\includegraphics[width=8cm]{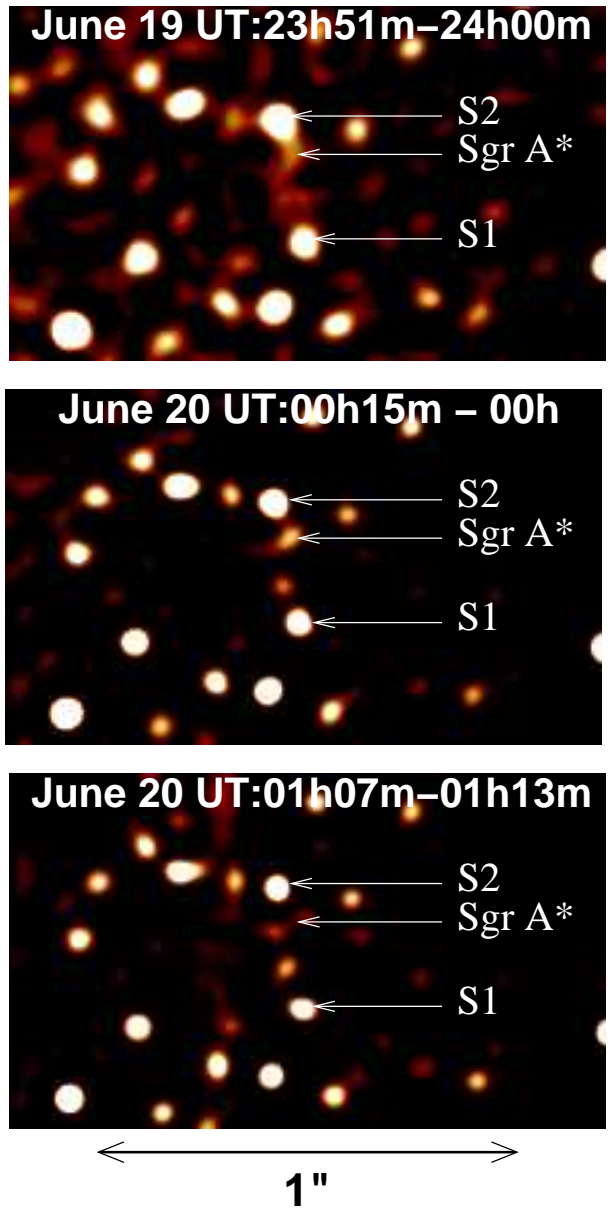}
\caption{\small Each panel shows the average of eight Lucy-Richardson
  deconvolved and beam restored images taken at different intervals of
  the measurements (see light curves in Figure~\ref{nircurves}). The
  VLT 8.2~m Unit Telescopes of the first image in each series are 
  indicated in the panels. 
  Top: Beginning of the observations. Middle: About 25~min
  after begin of the observations. Bottom: Around 80~min after begin
  of the observations.  One can also see how the image quality
  improved when comparing the upper and middle/lower panel images.
  Sgr~A* can be seen as a flaring source in the top and middle panels.
  The field of view is $1.43''$ (72 light days) times $0.85''$ (43
  light days).  }
\label{images}
\end{figure*}

The flaring of Sgr~A* can be seen in Figure~\ref{images}. In the top
panel we show the average of eight images at the beginning of the
observations. The middle panel shows the average of eight images that
were obtained about 25~min after the begin of the observations. Eight
images about 80~min into the observations are averaged in the bottom
panel. The VLT 8.2~m Unit Telescopes of the first image of each 
series are indicated in all
the panels. All images resulted after Lucy-Richardson deconvolution
and beam restoration. One can see how the AO correction improved
between images~1 and 2. Sgr~A* can be seen as a flaring source in the
first two images and is not visible in the last image.

The offset of the flaring source at the position of SgrA* from the
dynamical position of SgrA* - taking into account the orbit of S2
(Sch\"odel et al. 2003) - is 6$\pm$2~mas in R.A. and 12$\pm$4~mas in
Dec. and well within the values obtained for the previously detected
flares of SgrA* (Ghez et al. 2004, Genzel et al. 2003). The deviation
from the nominal position of Sgr~A* is most probably due to the
weakness of the flare and due to the proximity of faint stellar sources.

In parallel to the NIR observations, SgrA* was observed with \emph{Chandra}
using the imaging array of the Advanced
CCD Imaging Spectrometer (ACIS-I; Weisskopf et al., 2002) for
25.1~ks
on 19--20 June 2003 (UT).  The start and stop times are
listed in Table~\ref{log}.  The instrument was operated in
timed exposure mode with detectors I0--3 turned on.  The time between
CCD frames was 
3.141~s.  
The event data were telemetered in faint
format.

We reduced and analyzed the data using CIAO v2.3\footnote{Chandra
Interactive Analysis of Observations (CIAO),
http://cxc.harvard.edu/ciao} software with Chandra CALDB
v2.22\footnote{http://cxc.harvard.edu/caldb}.  Following
Baganoff et al. (2003), we reprocessed the level~1 data to remove the
0.25\arcsec\ randomization of event positions applied during standard
pipeline processing and to retain events flagged as possible
cosmic-ray afterglows, since the strong diffuse emission in the
Galactic Center causes the algorithm to flag a significant fraction of
genuine X-rays.  The data were filtered on the standard ASCA grades.
The background was stable throughout the observation, and there were
no gaps in the telemetry.

The X-ray and optical positions of three Tycho-2 sources were
correlated (H{\o}g 2000) to register the ACIS field on the Hipparcos
coordinate frame to an accuracy of 0.10\arcsec\ (on axis); then we
measured the position of the X-ray source at Sgr~A*.  The X-ray position 
[$\alpha$$_{J2000.0}$ = $17^{\mathrm h}45^{\mathrm m}40.030^{\mathrm s}$,
$\delta$$_{J2000.0}$ = $-29\arcdeg00\arcmin28.23\arcsec$] 
is consistent with the radio position of Sgr~A* (Reid et al. 1999) to 
within $0.18\arcsec\ \pm 0.18\arcsec$ ($1\,\sigma$).

We extracted counts within radii of 0.5\arcsec, 1.0\arcsec, and
1.5\arcsec\ around Sgr~A* in the 2--8 keV band.  Background counts
were extracted from an annulus around Sgr~A* with inner and outer
radii of 2\arcsec\ and 10\arcsec, respectively, excluding regions
around discrete sources and bright structures
(Baganoff et al. 2003).  The mean count rates within each radius
are listed in Table~\ref{tab:rates}.  The background rates have been
scaled to the area of the source region.  We note that the mean source
rate in the 1.5\arcsec\ aperture is consistent with the mean quiescent
source rates from previous observations 
(Baganoff et al. 2001, Baganoff et al. 2003).
The PSF encircled energy within each aperture increases from
$\approx50\%$ for the smallest radius to $\approx90\%$ for the
largest, while the estimated fraction of counts from the background
increases with radius from $\approx5\%$ to $\approx11\%$.  Thus, the
1.0\arcsec\ aperture provides the best compromise between maximizing
source signal and rejecting background.
\\
\\
We observed Sgr A* at 3.4 mm wavelength with the 
Berkeley-Illinois-Maryland Association (BIMA)
array (Welch et al. 1996) in its D configuration on 19 and 20 June 2003.  
Observations on both days were performed identically over the LST 
range 17.5 to 19.0 hours (07:47 to 09:17 UT on 19 June 2003).
These two observations occur approximately 9 hours before the start of
the Chandra observations and 4 hours after the end of the VLT 
observations, respectively.
This is 8-9 hours before and 7-9 hours after the end of the flare 
observed in the X-ray/NIR wavelength domain.
We observed the compact extragalactic source NRAO 530 (J1733-1303)  
three times
during the course of each observation.  For both sources, we
applied only the {\it a priori} amplitude calibration.
\\
\\
The D configuration is the most compact configuration, with
baseline lengths ranging from 1 to 9.5 kilolambda, corresponding to
a resolution of about 20 arcsec.  
The substantial
confusion from free-free emission on these baselines makes estimates of 
the absolute flux density and time variability during
these scans difficult.  
Comparison with past longer baseline 
observations allows us to determine
the contribution of the free-free emission and to estimate
the absolute flux density of Sgr A* to be $1.5 \pm 0.2$ Jy on 19 June 
2003 (Bower et al. 2001).
We are able to determine the difference in flux density between
the two epochs with much greater precision by differencing the two 
epochs.  We find
a marginally significant increase in the flux density of Sgr A* from 19 
June to 20 June of
$11 \pm 3$ \%.  The flux density of NRAO 530 changed by significantly 
less,
$-0.3 \pm 0.4$\%.


\section{Variability Analysis}

Figure~\ref{nircurves} and Figure~\ref{images} show the clear
detection of a decaying flare in the NIR emission of SgrA*.  This
flare is accompanied by an apparently simultaneous flare event in the
X-ray domain (see Figures~\ref{nircurves} and ~\ref{xray}).  The
following analysis consolidates the significance of the X-ray flare
and the correlation with the NIR flare event.

As a first step in our variability analysis of the X-ray data,
we constructed binned
light curves of the source intensity for each aperture using 10 minute
bins, then we tested each curve against the null hypothesis of a
constant count rate.  The $\chi^2$ statistic and degrees of freedom
for each fit are listed in Table~\ref{tab:rates}.  In each case, a
constant count rate yielded an acceptable fit to the data; for such
low count rates, however, $\chi^2$ analyses of binned light curves are
not particularly sensitive tests for variability.

Figure~\ref{xray} shows the source and background light curves for
the 1.0\arcsec\ aperture.  We note that the three highest points are
clustered within a 40 minute interval centered around 300 minutes
after the start of the \emph{Chandra} observation, and 
the first VLT/NACO image was taken 0.38 minutes before the midpoint 
of the highest X-ray data point.
The NIR light curve in Figure~1 shows clearly that Sgr A* was in a 
flaring state at this time.

To investigate this further, we applied a standard Kolmogorov-Smirnov
(K-S) test to the arrival times of the X-ray photons within each
aperture.  Multiple photons were occasionally detected in the same CCD
frame.  We corrected for this by redistributing their arrival times
over the frame interval using uniform random deviates.  The K-S test
revealed no evidence for significant variability (see
Table~\ref{tab:cumdist}).  The sensitivity of this test, however,
depends on the location of the maximum deviation along the
distribution.  Press et al. (1992) present a variant of the K-S test called
the Kuiper test that solves this problem by using the sum of the
maximum deviations of the observed cumulative distribution
\emph{above} and \emph{below} the theoretical distribution rather than
the maximum absolute deviation used in the K-S test.  The results of
the Kuiper test are shown in Table~\ref{tab:cumdist}.  The evidence
for variability in the 1.0\arcsec\ and 1.5\arcsec\ apertures, which
contain a higher fraction of background counts from diffuse emission,
was marginally significant: $\approx95\%$ confidence; while the
evidence for variability in the 0.5\arcsec\ aperture was highly
significant: 99.4\% confidence.  The results of the Kuiper test thus
suggest that Sgr A* varied in X-rays; even so, it gives us no
objective information about the location, duration, and amplitude of
the variability.

A flexible method for obtaining such information is the Bayesian
blocks algorithm of Scargle (1998), which uses the Poisson
distribution and Bayesian statistics to partition an interval of data
into piecewise constant segments or blocks.  Each block is modeled as
a Poisson process with constant intensity.  A revised version 
(Scargle et al. 2004) of the original method incorporates an algorithm for
finding the \emph{global}, \emph{optimal} partitioning of data on an
interval (Jackson et al., 2003).

The algorithm uses a geometric prior of the form $P(k) =
C\gamma^{-k}$, where $k$ is the number of change points, $P(k)$ is the
prior probability distribution on the number of change points, and
$\gamma$ is an adjustable parameter (Scargle et al., 2003, 2004).
Taking the logarithm of both sides yields the following contribution
to the log-posterior: $\log P(k) = \log C - k\log\gamma$.  This
additive term in the fitness function penalizes more complex models;
that is, models with more blocks.  Setting $\gamma = 1$ corresponds to
choosing a uniform prior, $P(k) = C$, where all values of $k$ are
equally likely.  The higher $\gamma$ is set, the harder it becomes for
the algorithm to add change points defining more blocks.

The value of $\gamma$ can be converted into a prior odds ratio, $P(k
\ge 1) / P(0)$, giving the ratio---before analyzing the
data---of the probability that the source varied at all to the
probability that it was constant.  This would equal $1 / (\gamma - 1)$
for an infinite geometric series, but $k$ can never be greater than
the number of detected photons, so the actual form of the odds ratio
is more complicated (J.~D.\ Scargle 2003, private communication).  In
practice, however, the series converges rapidly for $\gamma \gg 1$.

In Bayesian statistics, the posterior probability that the model is
correct given the data is proportional to the product of the prior
probability, expressing our knowledge or ignorance of the truth of the
model before analyzing the data, and the likelihood function, which is
the probability of the data given the model (Sivia 1996).  As noted
above, the prior odds ratio introduces a bias against more complex
models.  The preference of the data for the more complex model must be
compelling enough to overcome the prior bias before the algorithm will
introduce a new change point.  In other words, the ratio of the
likelihood functions for the data given the models must be greater
than the inverse of the prior odds ratio.

We applied the global Bayesian blocks algorithm to a binned light
curve of the total counts in the 1.0\arcsec\ aperture using 157.052
second bins (i.e., 50 times the interval between CCD frames).
Figure~\ref{blocks} shows the Bayesian blocks decomposition of
the light curve obtained with $\log\gamma = 3.350$, corresponding to a
prior odds ratio of $4.47 \times 10^{-4}$.  Two change points were
found 279.1 and 321.2 minutes into the observation indicating that
Sgr~A* flared in X-rays around midnight.  Similar results were
obtained for bin sizes ranging from 10 to 200 times the interval
between CCD frames.

The mean count rate and start and stop times for the three blocks are
presented in Table~\ref{tab:bblocks}.  The count rates during blocks 1
and 3 were fully consistent with each other.  The mean rate for these
two intervals combined was ($3.95 \pm 0.42$) $\times 10^{-3}$ counts
s$^{-1}$, whereas the mean rate during block 2 was $(11.63 \pm 2.59)
\times 10^{-3}$ counts s$^{-1}$: a difference of $(7.68 \pm 2.20)$
$\times 10^{-3}$ counts s$^{-1}$.

We performed Monte Carlo simulations to determine the posterior
probability of obtaining a block with a rate greater than or equal to
that of block 2.  Using the mean rate over the entire observation
(i.e., $4.40 \times 10^{-3}$ counts s$^{-1}$), we generated 100,000
simulated Poisson data sets with each set having the same number of
counts as the real data.  We then binned the counts into light curves
and ran the Bayesian blocks algorithm on each simulated light curve as
before.  The algorithm gave a false positive rate or posterior
probability of $7.7 \times 10^{-4}$, which is less than twice the
prior odds ratio estimated above.  Thus, the null hypothesis of a
constant rate is rejected with 99.923\% confidence, and we conclude
that Sgr~A* flared in X-rays for a period of about 42 minutes, which
is characteristic of both the X-ray and NIR flares detected in
previous observations (Baganoff et al., 2001, 2003; Goldwurm et al.,
2003; Porquet et al., 2003; Eckart et al. 2003, 
Genzel et al., 2003; Ghez et al., 2004).

\subsection{Properties of the Lightcurves}

\begin{figure*}
\centering
\includegraphics[width=12cm]{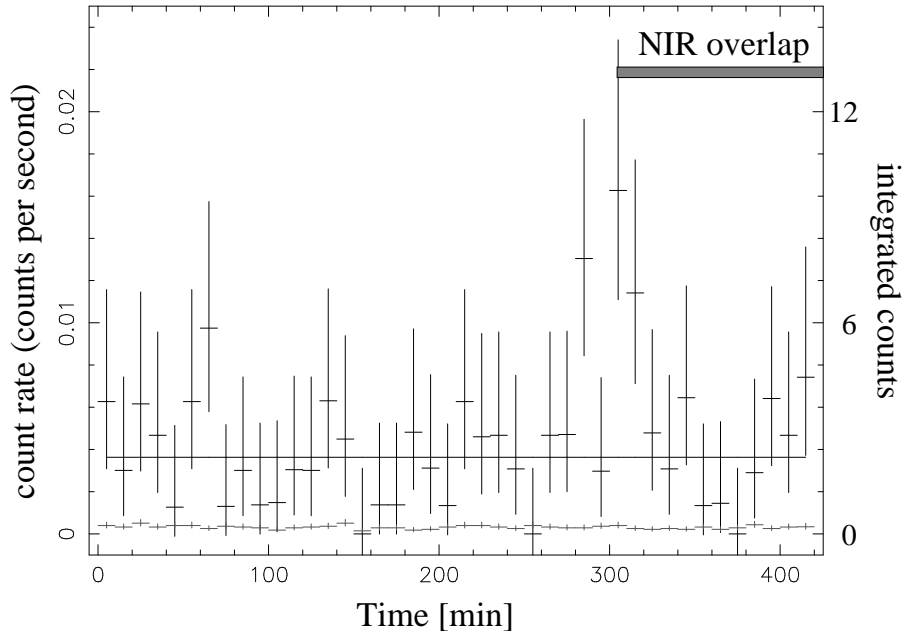}
\caption{\small 
X-ray light curve as observed by \emph{Chandra} in the
1.0\arcsec\ aperture.
The bin interval is 10 minutes. 
The solid straight line represents the X-ray IQ-state count rate.
The background (lower points) was steady throughout the 
observations which started on 19 June 2003 at 18:46:38 (UT).}
\label{xray}
\end{figure*}

From linear fits to the data in the rising and decaying flanks of the
X-ray and NIR flare including the measurement uncertainties
(Figures~\ref{nircurves} and \ref{xray}) we can estimate the times at
which the flare emission was negligible, i.e. equal to the IQ-state
emission.  The corresponding full width at zero power (FWZP) and start
and stop times (see Table~\ref{properties}) compare well with the
times of the change points and widths derived from the Bayesian blocks
routine.  They are earlier and later than the respective turn-on and
turn-off change points. Also an estimate of the FWHM
(Table~\ref{properties}) of the flare from Figure~\ref{xray} is
comparable to the flare duration derived from the Bayesian blocks
analysis.

Further support for the detection of a simultaneous flare at NIR and
X-ray wavelengths comes from a cross-correlation between the
variability data obtained in both domains.  The analysis was performed
on the measured NIR lightcurve (Figure~\ref{nircurves}, upper panel)
and on the X-ray data shown in Figure~\ref{xray}.  A constant flux
density contribution to SgrA* (see next section and Tab.1) was
subtracted before the analysis.  The cross-correlation was performed
by shifting the entire NIR data set over a range of -30 to +50
minutes with respect to its beginning at 19 June 23:51:15 UT.  We
cross-correlated only the data that overlap in time.  The result is
shown in Figure~\ref{correlation}.  With respect to the noise for
shifts of less than -20 and more than +20 minutes the
cross-correlation shows a $>$$5\sigma$ peak.  The graph shows a clear
maximum close to 0 minutes offset indicating that within the binning
sizes both data sets are well correlated.  
Although the IR data started 0.38 minutes before the midpoint of the
highest X-ray data point, a possible time lag between
the X-ray and NIR data of about 10 minutes (with the X-ray data leading)
is indicated. 
However, with the used binning sizes of 10 minutes (X-rays) and 
40 seconds (NIR) we therefore regard $\sim$15~minutes as a
conservative upper limit of any time lag between the NIR and X-ray
emission that may have been present on the {\it decaying} flanks of the
observed flare (Table~\ref{properties}).

In summary the statistical analysis of the combined X-ray and NIR data
shows that SgrA* underwent a significant flare event simultaneously in
both wavelength regimes.

\section{Flux Densities and Spectral Indices}

The IQ-state X-ray count rate in a 1.5'' radius aperture of
$5.3\pm0.5\times$10$^{-3}$~counts~s$^{-1}$ during the 
monitoring period is consistent with rates measured
during previous \emph{Chandra} observations (Baganoff et al. 2001, 2003).  
It corresponds to a
2 - 8 keV luminosity of 2.2$\times$10$^{33}$~erg/s or a flux density
of $0.015\mathrm{\mu}$Jy. The excess flux density observed during the
simultaneous flare
event was 0.039$\mu$Jy.  In total flux density this
is a factor of 3.6 higher and in excess flux density this is a factor
of 2.6 higher than the IQ-state.  This corresponds to a 2 - 8 keV
luminosity of about 6$\times$10$^{33}$~erg/s.  In the infrared the
$2.2\mathrm{\mu}$m flux density of the IQ-state Sgr~A* counterpart is
$1.9$~mJy and the excess flux density observed during the flare is
$3.7$~mJy (measured for the high S/N data points about 25~min after
the beginning of the observations).  
Since we may have passed the peak
of the NIR flare emission (see previous section and compare
Figures~\ref{nircurves} and \ref{xray}) this is in total flux density
at least a factor of 2.9 higher and in excess flux density this is at
least a factor of 1.9 higher than the IQ-state.

Hence, the spectral indices between the NIR regime (here at a
wavelength of $2.2 \mathrm{\mu}$m) and the X-ray domain (here centered 
approximately at an energy of 4~keV) of both the IQ- and the
flaring states are very similar, with $\alpha_{X/NIR}$$\sim$1.3.

For the IQ-state this spectral index is almost identical to 
$\alpha_{X/NIR}$$\le$1.36 as
calculated using the published flux density values in
Baganoff et al. (2001, 2003) and Genzel et al. (2003).  
However, with current data, it appears that the magnitude of
the fluctuations is substantially larger at X-ray 
(Baganoff et al. 2001, 2003; Porquet et al. 2003; Goldwurm et al. 2003, 
Eckart et al. 2003) than at IR wavelengths 
(Genzel et al. 2003, Ghez et al. 2004), so that the
spectral index could be smaller than about $\alpha_{X/NIR}$$\le$1.0
if we simply take the most extreme fluxes so far measured at
both wavelengths.

Our data also allow us to derive an estimate of the flux density
rise- and decay-rates (Tab.1). These rates are essential quantities
that describe possible combinations between time variations in the source 
geometry and the relevant energy release or dissipation processes.
The fractional rate of decay for both the X-ray and NIR
regimes is in agreement, within the measurement uncertainties.


\section{Physical Interpretation}

Current models that explain the SgrA* spectral energy distribution
invoke radiatively inefficient accretion flow models (RIAFs: Quataert
2003, Yuan et al. 2002, Yuan, Quataert, \& Narayan 2003, 2004,
including advection dominated accretion flows (ADAF): Narayan et
al. 1995, convection dominated accretion flows (CDAF): Ball et
al. 2001, Quataert \& Gruzinov 2000, Narayan et al. 2002, Igumenshchev
2002, advection-dominated inflow-outflow solution (ADIOS): Blandford
\& Begelman 1999), jet models (Markoff et al. 2001), and Bondi-Hoyle
models (Melia \& Falke 2001). 
Also combinations of models such as an accretion flow 
plus an outflow in form of a jet are considered 
(e.g. Yuan, Markoff, Falcke 2002).

Micro-lensing can most likely be excluded because of the short
duration of the events, their high frequency of occurrence, and the
shape of their light curves (see also Porquet et al., 2003; Ghez et
al., 2004; and Genzel et al., 2003).  The model in which stars
interact with an inactive, cold accretion disk (Nayakshin, Cuadra,
Sunyaev 2004, Nayakshin \& Sunyaev 2003) is also not a very
satisfactory explanation for the NIR flares.
To within a few
milliarcseconds the emission peaks of all NIR flares observed so far
(Genzel et al. 2003, Ghez et al. 2004) are located at the position of
the radio source SgrA*.  However, the most likely inner disk radius
required by the cold disk model is of the order of 10mas (Nayakshin \&
Sunyaev 2003), implying that all the flares did not occur within such
a disk - if it were present.  In addition, this model does not provide
an explanation for the $\sim$17 minute quasi-periodic fine structure
observed in 2 of the flares (Genzel et al. 2003). An analysis of the
two brightest X-ray flares also shows a temporal power spectrum that
is incompatible with the cold disk model (Aschenbach et al. 2004).

For the IQ-phase the theoretical models have to take into account that
the X-ray flux density is extended over the central 0.6 arcsecond
radius, while the flaring source is a point source 
(Baganoff et al. 2001, 2003).  For the flare
activity the short variation time
scale presents an additional complication to these models. Thermal
bremsstrahlung emission from the outer regions of an accretion flow
(R$>$10$^3$ R$_s$; see also Quataert 2003) would not be able to
explain the observed rapid variations of the X-ray flux density. Burst
models involving bremsstrahlung would require multiple components of
plasma at different temperatures because the NIR emission requires a
component of significantly lower temperature than the X-ray emission
(Genzel et al. 2003).  Pure synchrotron emission models require a high
energy cutoff in the electron energy distribution with large Lorentz
factors for the emitting electrons of $\gamma_e$$>$10$^5$ and magnetic
field strengths of the order of 10-100~G in order to explain the X-ray
emission. The correspondingly short cooling time scales of less than a
few hundred seconds would then require 
repeated injections or acceleration
of such energetic particles (Baganoff et al. 2001, Markoff et al. 2001,
Yuan, Quataert, \& Narayan 2004).

Our new simultaneous X-ray/NIR detection of the SgrA* counterpart
suggest that at least for the observed flare it is the same population
of electrons that is responsible for both the IR and the X-ray
emission, regardless of the emission mechanism.  While it is not yet
possible to completely rule out any of the proposed models, we find
that an attractive mechanism to explain the observed simultaneous
NIR/X-ray flare is the synchrotron self-Compton (SSC) process.  In
this model, the X-ray photons are produced by up-scattering of
millimeter or sub-millimeter photons.  
The (marginally) higher mm-flux density we observed after the flare may be
related to the preceding acivity at X-ray and NIR wavelengths.
The current data is not sufficient to constrain physical models
with a full coverage from the mm- to the X-ray domain.
The SSC process, however, directly couples both the NIR
and X-ray emission to the sub-mm-domain and would naturally result in
low upper limits on any measurable time lag between the NIR and X-ray
emission (see Tab.1 and previous section).
The short time scale of
the observed flux density variations indicates that the 
flare emission
originates from compact components within less than 10 Schwarzschild
radii R$_s$=2GM/c$^2$ with R$_s$=8.8$\times$10$^9$~$m$ for a
3$\times$10$^6$\solm ~black hole.  At a distance to the Galactic
Center of 8~kpc (Reid 1993, Eisenhauer et al. 2003) this corresponds
to an angular diameter of 6.8$\times$10$^{-3}$~mas.  For such a
component the inverse Compton scattered flux density will depend on
the Lorentz factor $\gamma_e$=(1-$\beta_e$$^2$)$^{-1/2}$ of to
relativistic electron distribution as well as 
the relativistic bulk motion of the emitting source with a Doppler
boosting factor $\delta$=$\Gamma$$^{-1}$(1-$\beta$cos$\phi$)$^{-1}$.
Here $\phi$ is the angle of the velocity vector to the line of sight,
$\beta$ the velocity $v$ in units of the speed of light $c$, and
Lorentz factor $\Gamma$=(1-$\beta$$^2$)$^{-1/2}$ for the bulk motion.
\\
\\ 
Doppler boosting will occur in models that involve relativistic
outflows or jets pointing towards the observer at a small angle to the line of
sight (e.g. Markoff et al. 2001). In the context of this jet model,
the emitting component would be located close to the jet base and
would have a size of a few $R_s$ or less.  
Bulk motion with
modest values of $\Gamma$$\le$2 can also occur in a
model in which matter is orbiting the black hole at small angles of
the orbital plane with respect to the observer's line of sight and
close to the innermost stable orbits (Bardeen, Press, \& Teukolsky
1972, Melia et al. 2001).
In this case the dominant component
responsible for a flare
would be the Doppler boosted material moving
toward the observer. The component size could then be of the order of
the Schwarzschild radius $R_s$.  In the following we
assume that the dominant component responsible for a flare
has a size
that is of the order of one to a few R$_s$.  

In such a model the rising and fading
times of the observed flares
(Baganoff et al. 2001; Genzel et
al. 2003) of a few 10 minutes (see Table~\ref{properties}) could
therefore be explained quite conveniently by Doppler boosting.  If the
matter or temperature distribution or the spatial distribution of
electrons in the high-energy tail of the electron energy distribution
over these orbits deviates significantly from a homogenous
distribution such a model would also be appropriate to explain the
quasi-periodic flux density variations observed in two NIR flares
(Genzel et al. 2003).  In this model more homogenous distributions
would also explain non-periodic variations over longer time scales,
i.e. the overall flare shape, by Doppler boosting of the approaching
side of the orbit.

We can compute the SSC spectrum produced by up-scattering
sub-mm-wavelength photons into the NIR and X-ray domain by using the
formalism given by Gould (1979) and Marscher (1983).  
Such a single SSC component model may be too simplistic
although it is considered
as a possiblity  in most of the recent modelling approaches.
For the single-component SSC model discussed in this paper, our data
suggests a NIR to X-ray spectral index of 1.3. This is inconsistent
with the X-ray spectral index of 0.3 observed by Chandra 
(Baganoff 2001, 2003), although it
is consistent with the spectral index of 1.5 +/- 0.3 reported by
Porquet et al. (2003).  
The spectral index, however, is a strong function of the
assumed ISM abundances and the calculation presented here should 
be taken as an illustrative example for a simple physical scenario.
No information on the in-band NIR- or X-ray spectral index is available 
for the present flare. 
In the NIR simultaneous recording of spectral information 
was not yet possible and in the X-ray domain the fluence of the reported 
flare is too small to derive reliable spectral index information.
We use the
notation by Marscher et al. (1983) and assume a synchrotron source of
angular extent $\theta$, that becomes optically thick at a frequency
$\nu_m$ with a flux density $S_m$, and has an optically thin spectral
index $\alpha$ following the law $S_{\nu}$$\propto$$\nu^{-\alpha}$.
This allows us to calculate the magnetic field strength $B$ and the
inverse Compton scattered flux density $S_{SSC}$ as a function of the
X-ray photon energy $E_{keV}$.  The synchrotron self-Compton spectrum
has the same spectral index as the synchrotron spectrum that is 
up-scattered 
i.e. $S_{SSC}$$\propto$$E_{keV}$$^{-\alpha}$, and is valid within the
limits $E_{min}$ and $E_{max}$ corresponding to the wavelengths
$\lambda_{min}$ and $\lambda_{max}$ (see Marscher et al. 1983 for
further details).

We find that Lorentz factors $\gamma_e$
for the emitting electrons of the order of 
a few thousand are required to produce a sufficient SSC flux in the
observed X-ray domain. While relativistic bulk motion 
is not a necessity to produce sufficient SSC flux density
we have used modest values for $\Gamma$ since they will occur
in case of relativistically orbiting gas as well as relativistic 
outflows - both of which are likely to be relevant in the case of 
SgrA*.

The flux densities $S_{2.2\mu m}$ and $S_{X-ray}$ of the
observed simultaneous flare
event can be explained very well for cases
in which the sub-millimeter emitting source component has a size of
the order of a few $R_s$ and a turnover
frequency $\nu_m$ of a few 100~GHz.  
For a relativistic bulk motion of
the emitting component with $\Gamma$=1.2-2 and $\delta$
ranging between 1.3 and 2.0 (i.e. angles $\phi$ between about
$10^{\circ}$ and $45^{\circ}$) the corresponding magnetic field
strengths are of the order of a few Gauss to about 20 Gauss,
which is within the range of magnetic fields expected for RIAF models
(e.g.  Markoff et al. 2001, Yuan, Quataert, Narayan 2003, 2004).
The resulting SSC spectrum is
defined over a contiguous wavelength range between the infrared to the
X-ray domain.  The low frequency cutoff of that up-scattered SSC
spectrum is at wavelengths between about  
100$\mathrm{\mu}$m and 1200$\mathrm{\mu}$m and the
high energy cutoff reaches energies between $\sim$100~keV to a few 1000~keV.
The required flux densities $S_m$ of the mm-/sub-mm-components 
are range between about 0.3 to 4 Janskys.
This simple SSC model would result in no time lag between the
NIR and X-ray emission - compatible with our limit on the time
lag for the decaying flank of the flare.

In case of $\nu_m$=70~GHz the required observable mm-flux 
density $S_m$ around 100 GHz is only of the order of 0.2 to 0.6 Jansky 
and compares very well with our flux density increase observed after 
the flare. 
This value is also well within the range of the observed 
variability of SgrA* in the mm-domain (Zhao et al. 2003).
For such a model, however, the required magnetic field strength 
is of the order of a few tenths of a Gauss and close to the minimum 
value required to have the cooling time of the flare less than the 
duration of the flare (Yuan, Quataert, Narayan 2003, 2004, Quataert 2003).
Higher values of $\nu_m$ are likely and would solve that problem.
Our time coverage, however, is not suited
to put tight constraints on the intrinsic emission process or on the role of
optical thickness effects as one would expect them e.g. 
in case of the adiabatic expansion of a synchrotron component (van der Laan 1966).


\section{Summary and Discussion}

We have presented the first successful simultaneous X-ray and NIR
detection of SgrA* in a flaring and the IQ-state.  The X-ray flare
lasted for about 42 minutes and began on 19 June 2003 at about 23:10
UT, more than 4 hours into the X-ray observation.  The peak of the
flare occurred 
less than 10 minutes prior to the time of the first VLT/NACO DDT image.  
The emission during the flare can successfully be
described by a SSC model in which the NIR and X-ray flux density
excess is produced by up-scattering sub-mm-wavelength photons into the
NIR and X-ray domain.

With respect to its FWZP duration of 55 - 115 minutes the flare
reported here compares well with other flares measured so far.
Baganoff et al. (2001), Eckart et al. (2003), and Porquet et
al. (2003) report on X-ray events of 45 to 170 minutes.  Ghez et
al. (2004) and Genzel et al. (2003) report on NIR flare events that
last 50 to 80 minutes, respectively.  Our newly detected flare event
is weaker than most others which have been reported and not
necessarily representative of the characteristics of the stronger
flares (factor of 50: Baganoff et al. 2001; factor of $>$100: Porquet
et al. 2003; Goldwurm et al. 2003).  However, during our 2002
\emph{Chandra} monitoring session we found that flares
that are a
factor of $>$10 stronger than the quiescent emission occur at a rate
of $0.53\pm0.27$ per day.  Weaker flares
are more frequent and our
newly detected flare event is probably more representative for those
weaker flares.

Although we have given preference to a simple SSC model in explaining
the observed simultaneous flare emission, flux density contributions
via other emission mechanisms may be of relevance.  Along with
enhanced electron heating leading to SSC flares, Markoff et al. (2001)
also suggested the possibility that the flares may result from
acceleration.  This could result from electrons which are energetic
enough to account for both the NIR and X-ray flares via direct
synchrotron radiation.  In fact even SSC models presented by Markoff
et al. (2001) and Yuan, Quataert, \& Narayan (2003) result in a
significant amount of direct synchrotron emission in the infrared (see
also synchrotron models in Yuan, Quataert, \& Narayan 2004).  The
different possible emission mechanisms may also be coupled in a more
complicated way. Ambient thermal electrons may be heated during a
flare and may produce excess sub-millimeter and infrared flux density.
This process could lead to correlated radio/NIR/X-ray variability
quite similar to to what is expected in SSC models (Yuan, Quataert, \&
Narayan 2003, 2004, Genzel et al. 2003, Ghez et al. 2004).  These
synchrotron/ inverse Compton models may also result in small (or no)
time lag between the NIR and X-ray emission - compatible with our
limit on the time lag for the decaying flank of the flare.  In such a
scenario it will be difficult to determine the relative importance and
flux density contributions of the different emission mechanisms by
variability.

Since the X-ray source responsible for most of the X-ray IQ-state flux
density of SgrA* is extended over a radius of about 0.6~arcsec
(Baganoff et al. 2003) its emission can most likely be ascribed to
bremsstrahlung from a thermal particle distribution.  While the
interim-quiescent NIR flux density probably can be attributed to a
fairly compact component (Genzel et al. 2003), it is currently not
clear how much of the quiescent X-ray flux density originates from
that compact region.  The extended X-ray component probably originates
from hot gas within the $\sim$1" Bondi radius that is associated with
the accretion flow (see Baganoff et al. 2003 and Quataert 2003 for detailed
discussions).

If thermal bremsstrahlung is an important mechanism to explain the
simultaneous NIR and the X-ray emission, then geometries are
conceivable in which the cooler $\ga$10$^3$~K plasma that gives rise
to the NIR emission is spatially offset from the hotter $\ga$10$^8$~K
plasma. One may also speculate about the possibility that the amount
of cooler plasma might be larger than the amount of hot plasma. These
facts could result in a situation in which the X-ray variations lead
the NIR-variations. On the other hand the cooling timescales of the
hotter plasma may be larger which would result in the opposite
behavior.  There are of course also models conceivable that involve
small source sizes or apparently cospatial distributions that would
result in no measurable time lag or even NIR and X-ray flare events
that need not always be correlated (see discussion by Yuan, Quataert,
\& Narayan 2004).

If we assume that the (marginally) higher mm-flux density after
the flare is related to the flare activity we observed in
the X-ray and NIR-domain, then there seems to be a 
correlation between flare activity in the 
mm/sub-mm and the X-ray domain in
three cases now: the original X-ray flare reported by
Baganoff et al. (2001; see also Zhao et al. 2004),
the activity around the largest X-ray flare reported by
Porquet et al. (2003) and Zhao et al. (2004),
and now the small X-ray flare we report on in this publication.

However, during an 8-day simultaneous observing campaign using the
Owens Valley mm-array at a wavelength of 3-mm and Chandra
in the X-ray domain (Mauerhan et al. 2004)
did not detect an obvious correspondence
of significant fluctuations in both wavelength domains.
The currently available data also indicate that the mm-activity 
may be a function 
of the X-ray flare magnitude. In the event reported by 
Porquet et al. (2003) and Zhao et al. (2004) the mm-flux density flared 
by 100\% of the continuum following the factor $160$ X-ray flare. 
Here we report a factor 2 to 3 X-ray 
flare for which we observed a no more than 10\% mm-flare following the
NIR/X-ray event.

Future observations will reveal possible relations
between different wavelength regimes. An important question is
whether individual mm-flare events are related to events in the
NIR or X-ray regime,  or whether there are in general more or
stronger flares per unit time,
when the average mm- or sub-mm flux density is higher.
Upcoming simultaneous monitoring programs from the radio to the
X-ray regime will be required to further investigate the physical
processes that give rise to the observed IQ-state and flare phenomena
associated with SgrA* at the position of the massive black hole at the
center of the Milky Way.


\begin{acknowledgements}
This work was supported in part by the Deutsche Forschungsgemeinschaft
(DFG) via grant SFB 494.  \emph{Chandra} research is supported by NASA grants
NAS8-00128, NAS8-38252 and GO2-3115B.  We are grateful to all members
of the NAOS/ CONICA team.  In particular we thank the ESO Director
General C. Cesarsky for supporting this project via Directors
Discretionary Time.  We are also greatful to N.~Ageorges and
L. Tacconi-Garman for discussions and support.

We thank J.\ Scargle for his explanations of the Bayesian blocks
method, and for his advice on numerical methods to determine the
posterior probability of features of interest in the resultant
segmented model.  We also thank M. Nowak for coding the Bayesian
blocks algorithm into S-Lang, and for discussing various aspects of
the numerical implementation of the method.

\end{acknowledgements}

\newpage


\begin{table}
\centering
\begin{tabular}{llccc}
\hline
                & & X-ray & NIR $2.2\mu m$ &  \\ \hline 
zero start time & June 19 UT & 23:00 - 23:20           & -             & \\
zero stop time  & June 20 UT & 00:15 - 01:05           & 00:35 - 00:55 & \\
FWZP            & min        & 55 - 115                & -             & \\
FWHM            & min        & 30 -  40                & -             & \\
decay rate      & nJy/min    & 0.75$^{+1.65}_{-0.45}$  &               & \\
                & mJy/min    &                         & 0.12$\pm$0.04 & \\
                &            &                         &               & \\
IQ-state           & $\mu$Jy & $0.015\pm0.004$ &              &  \\ 
flux density       &  mJy    &                 & $1.9\pm0.5$  &  \\ 
$\alpha_{X/NIR}$   &         &                 &              & $1.34\pm0.04$ \\ 
                   &         &                 &              &  \\
dereddened excess  & $\mu$Jy & $0.039\pm0.011$ &              &  \\
flux density       &  mJy    &                 & $3.72\pm0.35$  &  \\
$\alpha_{X/NIR}$   &         &                 &              & $1.29\pm0.04$ \\
                   &         &                 &              &  \\
time lag           &         &                 &              & $\le$15~min \\
\hline
\end{tabular}
\caption{X-ray and NIR Flare and IQ-State Properties. The start and
stop times - with an estimated error of 10-15 minutes - refer to those
times at which the rising and decaying flanks of the flare deviate
from the IQ-state (interim-quiescent) flux density.  The spectral
indices are defined via $S_{\nu}$$\propto$$\nu^{-\alpha}$.  The limit
on the time lag corresponds to the maximum lag between the decaying
flanks of the NIR and X-ray emission of the flare shown in
Figures~\ref{nircurves} and \ref{xray}.  The flare flux density is
given as peak flux density that was measured in excess of the IQ-state
flux density. The flare flux was measured for the high S/N
measurements taken about 25~min after the beginning of the
observations \label{properties}}
\end{table}


\begin{table}
\centering
\begin{tabular}{ccccc}
\hline
Telescope & Instrument & Energy/$\lambda$ & UT Start Time & UT Stop Time \\
\hline
\emph{Chandra} & ACIS-I & 2-8 keV & 19 JUN 2003 18:46:38  & 20 JUN 2003 01:45:13 \\
VLT UT4 & NACO & $2.18\mu$m & 19 JUN 2003 23:51:15 & 20 JUN 2003 03:53:58 \\
BIMA & -  & $3.4$mm         & 19 JUN 2003 07:47:00 & 19 JUN 2003 09:17:00 \\
BIMA & -  & $3.4$mm         & 20 JUN 2003 07:43:00 & 20 JUN 2003 09:13:00 \\
\hline
\end{tabular}
\caption{Observation Log.\label{log}}
\end{table}


\begin{table}[!h]
\begin{center}
\caption{X-ray Count Rates}
\label{tab:rates}
\vspace*{1ex}
\begin{tabular}{ccccc} \hline
\multicolumn{1}{c}{Extraction} &
\multicolumn{3}{c}{Count Rates\protect\footnotemark[1]} &
\multicolumn{1}{c}{$\chi^2$/d.o.f.\protect\footnotemark[2]} \\ \cline{2-4}
\multicolumn{1}{c}{Radius} &
\multicolumn{1}{c}{Total} &
\multicolumn{1}{c}{Source} &
\multicolumn{1}{c}{Background} &
\multicolumn{1}{c}{ } \\
\multicolumn{1}{c}{(\arcsec)} &
\multicolumn{1}{c}{($\times 10^{-3}$ cts s$^{-1}$)} &
\multicolumn{1}{c}{($\times 10^{-3}$ cts s$^{-1}$)} &
\multicolumn{1}{c}{($\times 10^{-3}$ cts s$^{-1}$)} &
\multicolumn{1}{c}{ } \\ \hline
0.5 & 1.61 $\pm$ 0.26 & 1.53 $\pm$ 0.10 & 0.080 $\pm$ 0.004 & \phn7.4/41 \\
1.0 & 4.72 $\pm$ 0.44 & 4.40 $\pm$ 0.13 & 0.320 $\pm$ 0.017 & 17.4/41 \\
1.5 & 7.02 $\pm$ 0.53 & 6.30 $\pm$ 0.15 & 0.719 $\pm$ 0.038 & 20.8/41 \\ \hline
\multicolumn{5}{l}{\footnotemark[1] ACIS-I count rate in 2--8 keV band.} \\
\multicolumn{5}{l}{\footnotemark[2] Fit of constant-rate model to source
  light curve using one-sided $1\,\sigma$ errors from Gehrels (1986).} \\
\end{tabular}
\end{center}
\end{table}

\begin{table}[!h]
\begin{center}
\caption{Cumulative Distribution Tests for X-ray Variability}
\label{tab:cumdist}
\vspace*{1ex}
\begin{tabular}{ccccccc} \hline
\multicolumn{1}{c}{Extraction} &
\multicolumn{1}{c}{Total} &
\multicolumn{1}{c}{K-S V} &
\multicolumn{1}{c}{P($>$ V)} &
\multicolumn{1}{c}{Kuiper D} &
\multicolumn{1}{c}{P($>$ D)} &
\multicolumn{1}{c}{Variable} \\
\multicolumn{1}{c}{Radius} &
\multicolumn{1}{c}{Counts } &
\multicolumn{1}{c}{Statistic} &
\multicolumn{1}{c}{ } &
\multicolumn{1}{c}{Statistic} &
\multicolumn{1}{c}{ } \\
\multicolumn{1}{c}{(\arcsec)} &
\multicolumn{1}{c}{(cts) } &
\multicolumn{1}{c}{ } &
\multicolumn{1}{c}{ } &
\multicolumn{1}{c}{ } &
\multicolumn{1}{c}{ } \\ \hline
0.5 & \phnn40 & 0.1654 & 0.2036 & 0.3198 & 0.0055 & Y \\
1.0 & \phn117 & 0.1170 & 0.0752 & 0.1592 & 0.0490 & ? \\
1.5 & \phn174 & 0.0838 & 0.1651 & 0.1348 & 0.0362 & ? \\ \hline
\end{tabular}
\end{center}
\end{table}

\begin{table}[!h]
\begin{center}
\caption{Bayesian Blocks Representation of X-ray Light Curve}
\label{tab:bblocks}
\vspace*{1ex}
\begin{tabular}{cccccc} \hline
\multicolumn{1}{c}{Block} &
\multicolumn{1}{c}{UT Start Time} &
\multicolumn{1}{c}{UT Stop Time} &
\multicolumn{1}{c}{Duration} &
\multicolumn{1}{c}{Counts\footnotemark[1]} &
\multicolumn{1}{c}{Count Rate\footnotemark[2]} \\
\multicolumn{1}{c}{ } &
\multicolumn{1}{c}{ } &
\multicolumn{1}{c}{ } &
\multicolumn{1}{c}{(s)} &
\multicolumn{1}{c}{(cts) } &
\multicolumn{1}{c}{($\times 10^{-3}$ cts s$^{-1}$)} \\ \hline
1     & 19 JUN 2003 18:46:37.85 & 19 JUN 2003 23:25:41.50 &    16743.65 & 65 & \phn3.93 $\pm$ 0.55 \\

2     & 19 JUN 2003 23:25:41.50 & 20 JUN 2003 00:07:48.84 & \phn2527.34 & 29 &    11.63 $\pm$ 2.59 \\

3     & 20 JUN 2003 00:07:48.84 & 20 JUN 2003 01:45:13.33 & \phn5844.48 & 23 & \phn3.99 $\pm$ 1.02 \\ \hline

\multicolumn{6}{l}{\footnotemark[1] Total ACIS-I counts within 1.0\arcsec\ radius of Sgr~A* in
                    2--8 keV band.} \\
\multicolumn{6}{l}{\footnotemark[2] Corrected for dead time between CCD frames; the correction
                   factor is 0.98693.} \\
\end{tabular}
\end{center}
\end{table}



\begin{figure*}
\centering
\includegraphics[width=12cm]{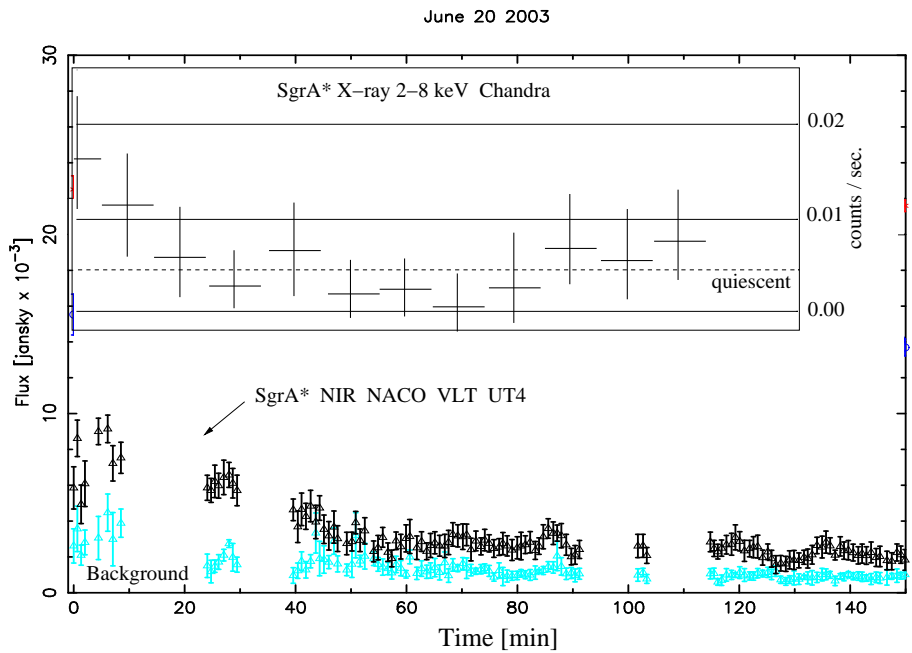}
\caption{\small The X-ray and NIR light curves plotted with a common
time axis.  See text and captions of previous figures.  Straight solid
lines in the inserted box represent the 0.00, 0.01, and 0.02 counts
per second levels.  The straight dashed line represent the X-ray
IQ-state flux density level.  The NIR observations started on 19 June
2003 at 23:51:15 (UT); see caption of Figure~\ref{nircurves}. }
The NIR data started 0.38 minutes before the midpoint of the
highest X-ray data point.
\label{lightcurves}
\end{figure*}


\begin{figure*}
\centering \includegraphics[angle=270,width=12cm]{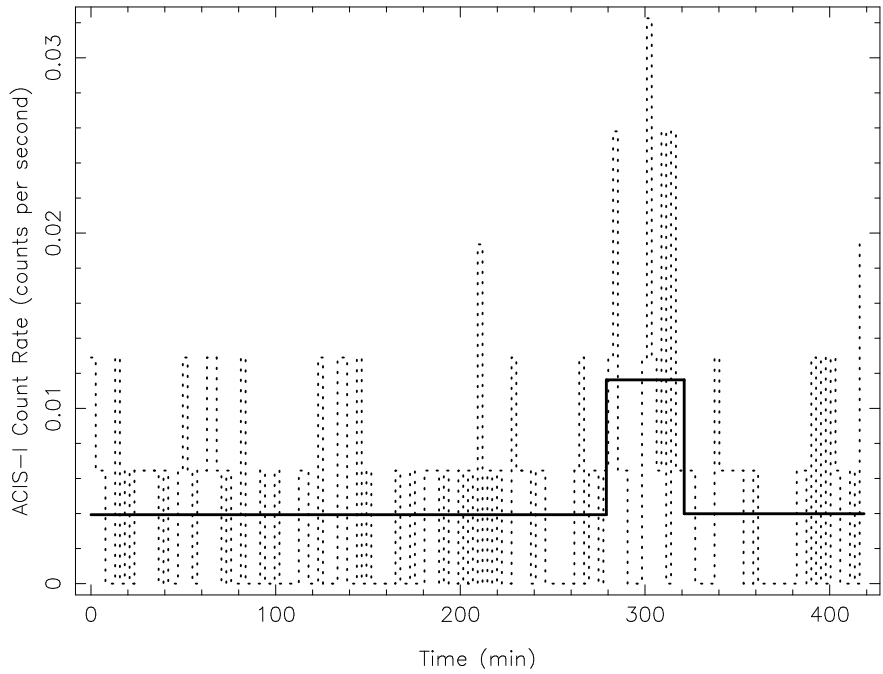}
\caption{\small 
Bayesian blocks representation (solid line) of the X-ray light
curve (dashed line) using 157.052 second bins. Two change points
detected with 99.927\% confidence indicate a flare
event around midnight during 
the interval 279 to 321 minutes into the
observation.
The time series starts on 19 June 2003 at 18:46:38 (UT).
}
\label{blocks}
\end{figure*}


\begin{figure*}
\centering \includegraphics[width=12cm]{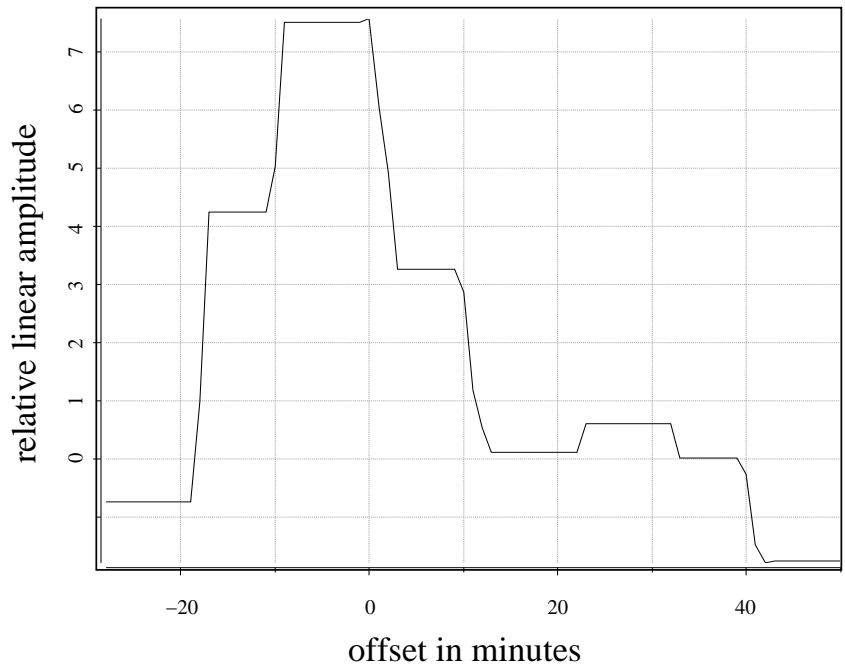}
\caption{\small Cross-correlation between the NIR data (40 second bins;
20 seconds integration time per image) and the X-ray data (10 minute bins).  
At a $>$ 5$\sigma$ level both light curves indicate a simultaneous flare
event around midnight corresponding to a time delay of less than 15
minutes (see text). 
 In Figure~\ref{correlation} we cross-correlated
only the flare data that overlap in time.  
The offsets are given in minutes
with respect to the beginning of the NIR data at 19 June 23:51:15 UT.
The plot is limited by this start time to the left.}
\label{correlation}
\end{figure*}


\end{document}